          [2001/04/25 1.1 (PWD)]
\documentclass[a4paper,twocolumn]{esapub2005} 
\usepackage{times}
\usepackage{natbib}
\usepackage{graphicx}

\title{INTEGRAL and RXTE observations of the fast X-ray transient XTEJ1901+014}
\author{D.I.Karasev}
\author{A.A.Lutovinov}
\author{S.A.Grebenev}
\affil{Space Research Institute, Profsoyuznaya str. 84/32, Moscow 117997, Russia}

\begin{document}


\def\arcmin{$^{\prime\, } $}

\maketitle

\begin{abstract}
We present results of spectral and timing analysis of the fast X-ray transient XTE J1901+014 based on data of the RXTE and INTEGRAL observatories. With the INTEGRAL/ISGRI the source was detected at a significance level of 20$\sigma$ with the persistent flux of $\sim$2.7 mCrab in a 17-100 keV energy band in 2003-2004 (during long observations of the Sagittarius arm region). We added the  RXTE/PCA (3-20 keV) data obtained in 1998 to the INTEGRAL/ISGRI data to build the broadband spectrum of the source in a quiescent state. It was found that the spectrum can be well approximated by a simple powerlaw with a photon index  of $\sim$2.15. From timing analysis we found short time scale aperiodic variations which can be connected with instabilities in the accretion flow. 
\end{abstract}


\section{Introduction}

The fast X-ray transient source XTE J1901+014 was discovered [4] by
the all-sky monitor ASM on board the RXTE observatory during the
powerful outburst on April 6, 2002 lasted from 3 min to 3.15 hours and
reached the peak flux $\sim$0.9 Crab in the 1.5-12 keV energy band
(Fig.1, right panel). The source position was determined as RA =
19$^h$ 01$^m$ 45$^s$.95, DEC = +1 24\arcmin 15.7\arcmin\arcmin (J2000;
3\arcmin uncertainty). The analysis of the archival ASM data [5]
revealed a previous outburst from the same position on June 21,
1997. This outburst was longer than 6 min and shorter than 8 hr, with
a peak flux of $\sim$0.4 Crab (Fig. 1, left panel). The obtained
information about XTE J1901+014 was not enough to make any confident
conclusions about its nature, but it was noted that the time scale of
this flare is similar to those of such events observed from the black
hole binary V4641 Sgr.

In this report we briefly present results of observations of
XTEJ1901+014 with the INTEGRAL and RXTE observatories. More detail
analysis will be presented separately (see [2]).

\section{Outburst and quescent state}

\begin{figure}[b]
\centering
\includegraphics[width=1\linewidth]{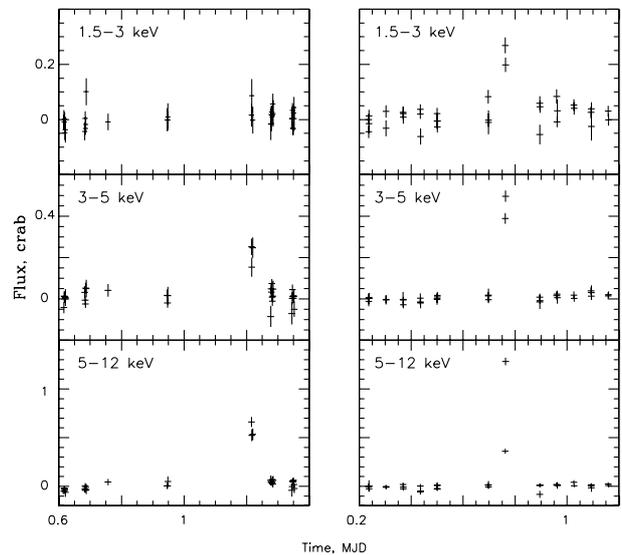}
\caption{Lightcurves (1.5-12 keV) of XTE J1901+014 measured by the RXTE/ASM during outbursts :  in June, 1997 (left pannel: 0 corresponds to UT 20/06/97 00:00:00); in April, 2002 (right panel: 0 corresponds to UT 06/04/02 00:00:00). One dwell is 90 s.}
\end{figure}

During the outburst in July 1997 the source flux in the 1.5 - 3 keV energy band did not exceed the background level whereas in the harder energy bands, 3-5 keV and 5-12 keV, it reached  $\sim$0.13 Crab and  $\sim$0.7 Crab, respectively.
 During the outburst in April 2002 the peak fluxes in these three bands were detected at the levels of $\sim$0.8, $\sim$1.1  and $\sim$1.2 Crab, respectively. Thus both observed outbursts were hard.

We analysed RXTE/ASM archive data from Junuary, 1996 to July, 2006 and could not find other such powerful outbursts from the source.
\begin{figure*}
\centering
\includegraphics[ width=18cm,bb=13 12 1300 520,clip]{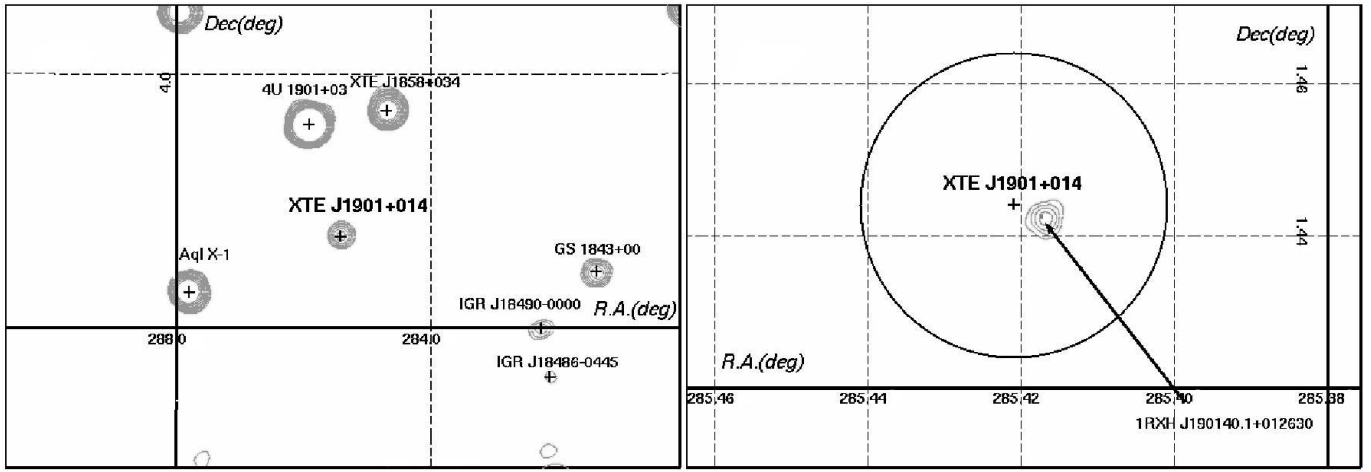}

\caption{Maps of the region of the sky that include the X-ray source XTE J1901+014 obtained by INTEGRAL/ISGRI (17-100 keV) (left panel) and 1RXH J190140.1+012630 obtained by ROSAT/HRI (0.1-2.4 keV) (right panel),  with the INTEGRAL/ISGRI error box for XTEJ1901+014 (a circle corresponds to a radius ~1.2\arcmin).}
\end{figure*}

XTEJ1901+014 was detected in the quiescent state (outside of
outbursts) by both the spectrometer RXTE/PCA in September, 1998 and
April, 2002, with the full exposure $\sim$1650 s and an average 3-100
keV flux of $\sim$2.8 mCrab (is was the same in different years) and
the detector INTEGRAL/ISGRI in 2003 - 2004 see above with an average
flux of $\sim$2.7 mCrab in the 17-100 keV energy band.

Some aperiodic variability of the source flux was detected in all RXTE
observations. We found a number of small flares with a duration of 40
- 60 s and a maximal flux of $\sim$6 - 8 mCrab. The origin of such
variability is most likely connected with a nonsteady accretion.

\section{Refinement of the position}

Analysis of the ROSAT All-Sky Survey Source Catalogue, has shown that
the source 1RXS J190141.0+012618 is located in the RXTE/ASM error box
(3\arcmin) of XTE J1901+014. During the pointed ROSAT/HRI observation
performed on October 3, 1994, the source was also detected, its
position was refined and the source was renamed as 1RXH
J190140.1+012630 [7].

Using the INTEGRAL/ISGRI data we improved an accuracy of the XTE
J1901+014 localization to $\sim$1.2\arcmin. As it clearly seen from
Fig.2 the ROSAT source 1RXH J190140.1+012630 confidently the distance
between positions of XTE J1901+014 and 1RXH J190140.1+012630 is about
0.3\arcmin) falls into the INTEGRAL/ISGRI error box for XTEJ1901+014,
that points that XTE J1901+014 and 1RXH J190140.1+012630 are the same
source.

\section{Spectral analysis}

We have only very poor information of the source spectral evolution
during the outbursts (see below), but can precisely reproduce its
spectrum in the quiescent state.

\subsection{Quiescent state}

To obtain the broadband spectrum of the source in the quiescent state
we used RXTE/PCA data in the 3 - 20 keV energy band and INTEGRAL/ISGRI
data in the hard energy band ($>$20 keV) analysis. It is important to
note, that the PCA/RXTE observations were performed in 1998, 2002 and
the ISGRI/INTEGRAL ones - in 2003-2004. Thus our spectral
reconstruction is correct in the suggestion that the spectrum shape of
the source does not change during this time interval.

The broadband (3-100 keV) spectrum of XTEJ1901+014 was approximated by
a simple power law model with the interstellar absorption which value
was fixed at N$_H$ = $0.7\times10^{22}$ atom/cm$^2$ that is typical
for this direction to the sky (it was evaluated from the N$_H$
map). The best-fit photon index is $\Gamma$=2.15 $\pm$ 0.03 (Fig. 3).

We analysed small short flares registered by RXTE/PCA from the source
(see above) and found that the source spectral shape did not changed
during the flares.

XTEJ1901+014 is located near the Galactic plane (l = 35.38 deg, b
=-1.62 deg), thus the Galactic Ridge emission could strongly affect
the result of spectral measurements with RXTE/PCA [3]. In this report
the spectrum and lightcurves of XTEJ1901+014 were obtained taking into
account this contamination. In order to estimate the Galactic ridge
emission intensity we used the data obtaned during pointing
observations of nearby transient sources performed during their
"turned off" state.

In particular we used pointing data of GS 1843-02 (l $\simeq$31 deg, b
$\simeq$-0.5 deg) observations, that is the nearest transient X-ray
pulsar to obtain the Galactic ridge spectrum at its position for
XTEJ1901+014. The analysis of this data allows us to obtain the
Galactic ridge spectrum near GS 1843-02. Due to the nature of the
Galactic ridge emission its spectrum has the same form in different
regions of the sky with -5 deg $<$ b $<$ +5 deg [3]. Therefore we can
just renormalize this spectrum (using the scan data), to get the
Galactic ridge spectrum at XTEJ1901+014 position.

The importance of accounting the Galactic ridge emission is demonstrated by Fig.4, where the total PCA/RXTE spectrum is shown along with the Galactic ridge and source true spectra.
\begin{figure}[t]
\centering
 \includegraphics[width=8cm,bb=30 200 580 710,clip]{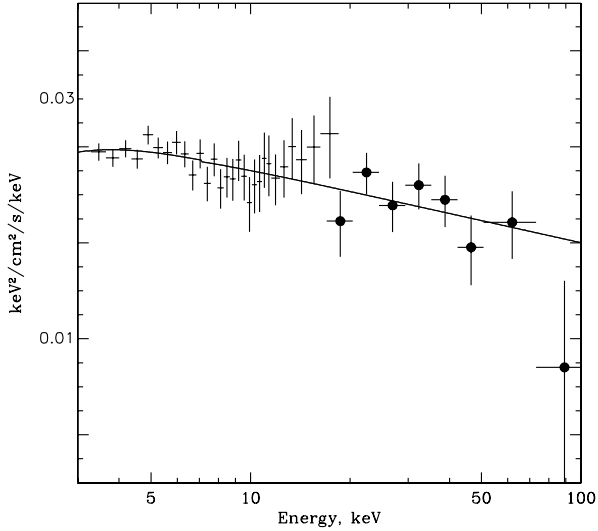}
\caption{Broadband energy spectrum (3-100 keV) of XTEJ1901+014 obtained by RXTE/PCA and INTEGRAL/ISGRI.}
\end{figure}

\subsection{The outbursts}

However using two energy bands of RXTE/ASM (3-5 and 5-12 keV) it is possible to roughly estimate evolution of  the photon index during the outbursts. According to [6] the photon index $\Gamma_{ASM}$ can be expressed as:
\begin{equation}
\Gamma_{ASM} = 1.499\times R + 0.698	
\end{equation}
where R - the relation between count rates in  the 3-5 keV and  5-12 keV energy bands. Note that this equation was obtained for sources with powerlaw spectra. 

For the outburst in June, 1997, we found that $\Gamma_{ASM}$ was not changed significantly; for the outburst in April, 2002 - it changed from 2.4 +/- 0.1  to 1.4 +/- 0.1. 
\begin{figure}[t]
\centering
 \includegraphics[width=9cm]{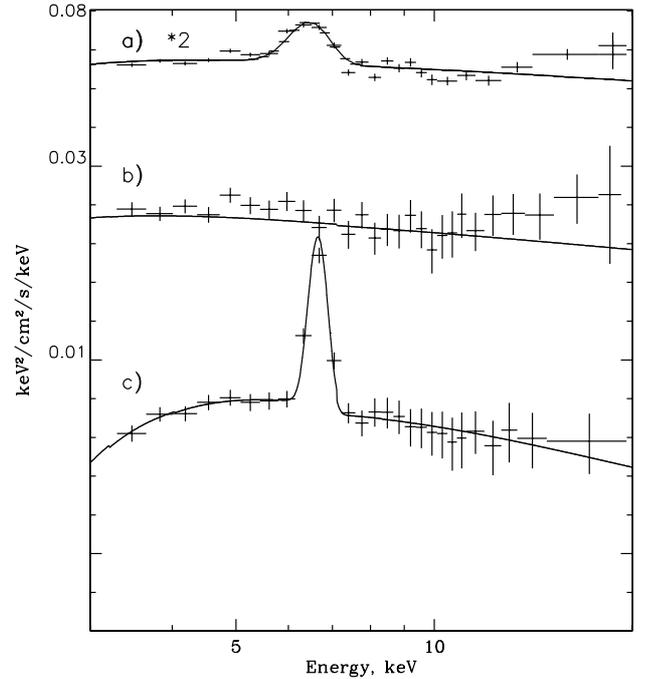}

\caption{ a) The rough spectrum measured by RXTE/PCA from the XTEJ1901+014 position (scaled by a factor of 2 for clarity); b) Reconstructed true spectrum of XTEJ1901+014  in the 3 - 20 keV energy range ;  c) The Galactic ridge spectrum of the XTEJ1901+014 position according to our estimate.}
\end{figure}

\section{Summary}

\begin{itemize}
\item The broadband spectrum of  XTEJ1901+014 in the quiescent state was obtained and investigated for the first time. It can be approximated by a simple power law model with the photon index of $\sim$2.15 without any high energy exponential cutoff. 
\item The powerful outbursts of the source in 1997 and 2002 are not the 1st type X-ray bursts because they become harder with time, resembling to the well-known outburst of V4641Sgr or outbursts from SAXJ1818.6-1703 [1].
\item A number of small short flares were detected from the source during pointed RXTE/PCA observations.
\item The accuracy of the XTEJ1901+014 localization was improved from 3\arcmin to $\sim$1.2\arcmin that strengthens the association of the source with the ROSAT soft source 1RXH J190140.1+012630.
\end{itemize}
Summarizing all the above we can suppose that XTEJ1901+014 belongs to the class of fast X-ray transient sources, but for the final answer on its origin, more observations at different wavelengths are necessary.

\section{ACKNOWLEDGEMENTS}

The authors thank to E.Churazov for the developing of the methods of
the analysis of the IBIS data and software. Authors also thank
M.Revnivtsev for the help with the analysis of RXTE data and a
discussion of the results obtained.  This work was supported by the
Russian Foundation for Basic Research (projects no.05-02-17454 and
02-04-17276), the Russian Academy of Sciences (The Origins and
evolution of stars and galaxies program) and grant of President of RF
(NSh- 1100.2006.2). AL acknowledges financial support from the Russian
Science Support Foundation.

\end{document}